# The Evolution of EU Business Cycle Synchronisation 1981 - 2007

**Paul Ormerod (pormerod@volterra.co.uk)**

**Volterra Consulting Ltd**

135c Sheen Lane
London SW14 8AE
UK

*Abstract*

*Most of the analytical techniques used in the business cycle convergence literature rely upon the estimation of an empirical correlation matrix of time series data of macroeconomic aggregates in the various countries, real GDP usually being the key variable. However due to the finite size of both the number of economies and the number of observations, a reliable determination of the correlation matrix may prove to be problematic. The structure of the correlation matrix may be dominated by noise rather than by true information.*

*Random matrix theory was developed in physics to overcome this problem, and to enable true information in a matrix to be distinguished from noise. It has been successfully applied in the analysis of financial data. The largest eigenvalue of the correlation matrix informs us directly about the degree to which the movements of the economies are genuinely correlated.*

*To follow the evolution of the degree of business cycle convergence over time we may analyse how the largest eigenvalue of the correlation matrix evolves temporally. The analysis is undertaken with a fixed window of data. Within this window the spectral properties of the correlation matrix formed from this data set are calculated. This window is then advanced by one period and the maximum eigenvalue noted for each period.*

*This paper applies the techniques to quarterly real GDP data 1981Q1 – 2008Q1 for the main EU economies, Germany, France, Italy, UK, Netherlands, Belgium and Spain, along with the US as a comparator. It extends previous results reported in physics journals (Ormerod and Mounfield (2002), Ormerod (2005)).*

*For the core EU countries, France, Germany, Italy, Spain, Belgium and the Netherlands, business cycles have shown strong synchronisation over the whole of the 1981-2008 period. The United Kingdom and the United States are considerably more synchronised with each other than they are with the main EU economies*

1. **Introduction**

This paper examines the extent to which the business cycles of the main EU economies have been in synchronisation over the 1981 - 2007 period, and how this has altered over this period.

I examine the performance of the EU 'core', the large economies of France, Italy and Germany plus Belgium and the Netherlands, which were founder members of both the EU itself and of the Euro, and the core plus the large economy of Spain, which did not join the EU until 1982 but which was a founder member of the Euro. This is contrasted with the core plus the UK, which whilst a member of the EU since 1973 has not joined the Euro and has been consistently the least supportive of ideas of further European integration. The United States is also included as a comparator.

I use the technique of random matrix theory (Mehta 1991) to analyse the correlations between the growth rates of the economies over time. Section 2 discusses the relevance of this theory, and section 3 sets out the empirical results.

2. **Random matrix theory**

Quarterly data exists for the main EU economies over the past thirty years or so for the level of real output in the economy (GDP). We can therefore calculate annual growth rates quarter-by-quarter. The correlations between these growth rates for the various economies will inform us about the extent to which their business cycles are in synchronisation.

In other words, the degree of synchronisation of the business cycles may be quantified by calculation of the correlation matrix of the matrix of observations formed from the time series of GDP growth for each economy.

If $\underline{\underline{M}}$ is an N x T rectangular matrix (T observations of the GDP growth of the N economies) and $\underline{\underline{M}}^T$ is its transpose, the correlation matrix $\underline{\underline{C}}$ as defined below is an N x N square matrix

$$\underline{\underline{C}} = \frac{1}{T} \underline{\underline{M}}\, \underline{\underline{M}}^T$$

However due to the finite size of N (which corresponds to the number of economies) and T (which is the number of observations of GDP) then a reliable determination of the correlation matrix may prove to be problematic. The structure of the correlation matrix may be dominated by noise rather than by true information.

In order to assess the degree to which an empirical correlation matrix is noise dominated we can compare the eigenspectra properties of the empirical matrix with the theoretical eigenspectra properties of a random matrix. Undertaking this analysis will identify those eigenstates of the empirical matrix who contain genuine information content. The remaining eigenstates will be noise dominated and hence unstable over time. This technique has been applied by many researchers to financial market data (for example, Mantegna et al 1999, Laloux et al 1999, Plerou et al 1999, Plerou 2000, Bouchaud et al 2000, Drozdz et al 2001).

For a scaled random matrix **X** of dimension N x T, (i.e where all the elements of the matrix are drawn at random and then the matrix is scaled so that each column has mean zero and variance one), then the distribution of the eigenvalues of the correlation matrix of **X** is known in the limit T, N $\rightarrow \infty$ with Q = T/N $\geq$ 1 fixed (Sengupta et al 1999). The density of the eigenvalues of the correlation matrix, $\lambda$, is given by:

$$\rho(\lambda) = \frac{Q}{2\pi} \frac{\sqrt{(\lambda_{max} - \lambda)(\lambda - \lambda_{min})}}{\lambda} \qquad \text{for } \lambda \in [\lambda_{min}, \lambda_{max}]$$

and zero otherwise, where $\lambda_{max} = \sigma^2 (1 + 1/\sqrt{Q})^2$ and $\lambda_{min} = \sigma^2 (1 - 1/\sqrt{Q})^2$ (in this case $\sigma^2 = 1$ by construction).

The eigenvalue distribution of the correlation matrices of matrices of actual data can be compared to this distribution and thus, in theory, if the distribution of eigenvalues of an empirically formed matrix differs from the above distribution, then that matrix will not have random elements. In other words, there will be structure present in the correlation matrix.

To analyse the structure of eigenvectors lying outside of the noisy sub-space band the Inverse Participation Ratio (IPR) may be calculated. The IPR is commonly utilised in localisation theory to quantify the contribution of the different components of an eigenvector to the magnitude of that eigenvector (thus determining if an eigenstate is localised or extended) (Plerou et al 1999).

Component $i$ of an eigenvector $v_i^\alpha$ corresponds to the contribution of time series $i$ to that eigenvector. That is to say, in this context, it corresponds to the contribution of economy $i$ to eigenvector $\alpha$. In order to quantify this we define the IPR for eigenvector $\alpha$ to be

$$I^\alpha = \sum_{i=1}^{N} (v_i^\alpha)^4$$

Hence an eigenvector with identical components $v_i^\alpha = 1/\sqrt{N}$ will have $I^\alpha = 1/N$ and an eigenvector with one non-zero component will have $I^\alpha = 1$. Therefore the inverse participation ratio is the reciprocal of the number of eigenvector components significantly different from zero (i.e. the number of economies contributing to that eigenvector).

## 3.  The data and the results

Quarterly levels of real GDP over the period 1980Q1 – 2008Q1 are available from the OECD database for the largest EU economies, France, Germany, Italy[1], Spain, UK, Belgium and the Netherlands.  All of these with the except of the UK are of course members of the Euro zone, and for purposes of description will be referred to as the 'core EU' economies

I analyse the correlation matrix of real GDP growth rates for various permutations of these economies.  In addition, the United States is included as a comparator.

For the data set of all eight economies as a whole, the eigenvalues are in the range 0.19 to 4.33.  The theoretical range of the eigenvalues of a random matrix of the same dimension is 0.53 to 1.61.  These results indicate the presence of a large amount of true information in the correlation matrix.

In terms of those eigenvalues which lie outside the noisy sub-space band the most important from a macroeconomic perspective is the largest eigenvalue. The application of these techniques to equities traded in financial markets have demonstrated that this eigenmode corresponds to the 'market' eigenmode (e.g. Gopikrishnan et al, 2000). In this context the largest eigenvalue will inform us as to the degree to which the movements of the EU economies are correlated.

The contribution which each of the economies makes to eigenvector 1 can be seen from calculating the IPR.  The value of the IPR is 7.15.

The trace of the correlation matrix is conserved, and is equal to the number of independent variables for which time series are analysed. That is, for the eight economies considered together, the trace is equal to 8 (since there are 8 time series). The closer the 'market' eigenmode (i.e. eigenmode 1) is to this value the more information is contained within this mode i.e. the more correlated the movements of GDP. The analysis therefore suggests the existence of a distinct international business cycle between these economies.

The market eigenmode corresponds to the largest eigenvalue. The degree of information contained within this eigenmode, expressed as a percentage, is therefore $100\lambda_{max}/N$. To follow the evolution of the degree of business cycle convergence over time we may analyse how this quantity evolves temporally. The analysis is undertaken with a fixed window of data. Within this window the spectral properties of the correlation matrix formed from this data set are calculated. In particular the maximum eigenvalue is calculated. This window is then advanced by one period and the maximum eigenvalue noted for each period.

---

[1] At the time of writing data for Italy was only available to 2007Q3, estimates for 2007Q4 and 2008Q1 were used

The choice of an appropriate window to span the periodicity of what constitutes the business cycle is not completely straightforward. Business activity is influenced by a very large number of events, and these events may be very diverse in character and scope. Individual cycles therefore vary both in terms of amplitude and period. This lack of regularity may be analysed formally using random matrix techniques (Ormerod and Mounfield 2000). The evidence for the existence of a business cycle at all relies more upon factors such as the fact that output changes in different sectors of an economy tend to move together (Lucas 1977) than upon regularities in either amplitude or period of the economy as a whole.

A major study of the US economy (Burns and Mitchell 1946) many years ago concluded that the period ranged from some two to twelve years, a range which still commands broad assent amongst economists. Analysing the data in the frequency domain, using the command 'spec.pgram' in the statistical package S-Plus[2], the spectrum of the US data is concentrated in the range 4 to 9 years. This compares with an estimate of 2 to 7 years obtained by Cogley and Nason (1995)[3], although in each case the degree of concentration is weak. For the UK, the estimated range is 7 to 12 years and for Germany, for example, 6 to 12 years. Again, the concentration at these frequencies is only weakly determined.

On the basis of the above, results are presented using a window of 8 years, and are set out in Figure 1. Each window contains 32 quarterly observations, and so there are 77 windows in total. The period 1981Q1 - 1988Q4 corresponds to the first data point in Figure 1, 1981Q2 - 1989Q1 to the second, and so on through to 2000Q2 – 2008Q1.

---

[2] See P. Bloomfield, *Fourier Analysis of Time Series: An Introduction.* Wiley, New York, 1976 and the chapter 'Analyzing Time Series' of the *S-PLUS Guide to Statistical and Mathematical Analysis* for specific details

[3] T. Cogley and JM Nason, 'Output dynamics in real business cycle models', *American Economic Review*, 85,492-511, 1995

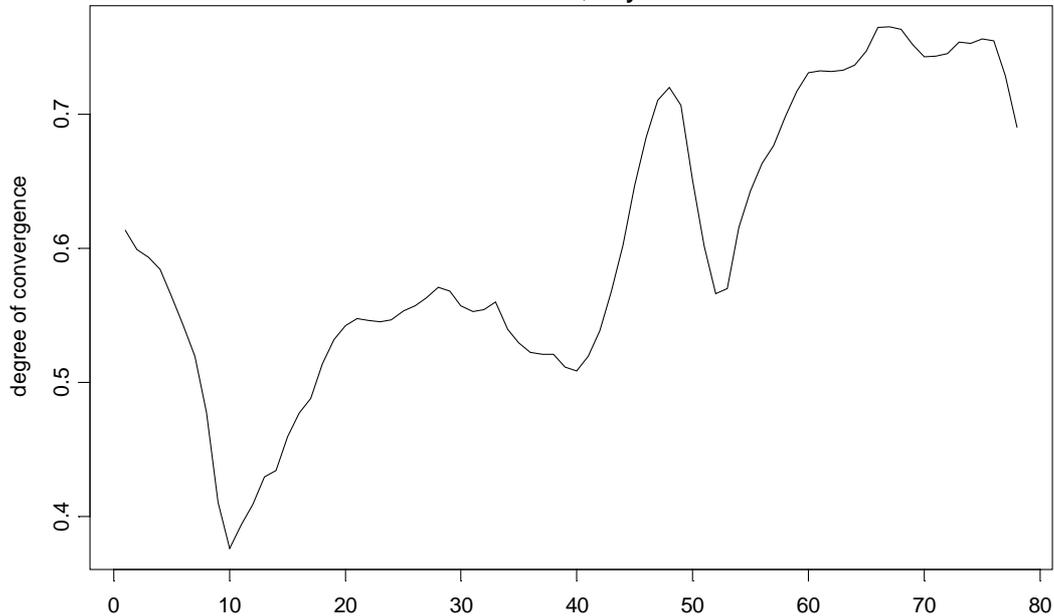

**Figure 1**
*The temporal evolution of the degree of information content in the maximum eigenvalue of the empirical correlation matrix formed from the time series of quarterly GDP growth for the economies of France, Germany, Italy, Belgium, Netherlands, Spain, UK and US. Each window of data spans 32 quarterly observations. The period 1981Q1 - 1988Q4 corresponds to the first data point in Figure 1, 1981Q2 - 1989Q1 to the second, and so on through to 2000Q2 - 2008Q1.*

Even in the early part of the period, the 'market' eigenvalue took up some 60 per cent of the total of the eigenvalues, indicating a strong degree of convergence of the business cycles of the relevant economies. There was a temporary reduction of convergence around the time of German re-unification in the early 1990s, but the economies have re-converged and by the 2000-2008 period, the principal eigenvalue accounted for nearly 80 per cent of the total information content within the correlation matrix, indicating a movement towards even greater convergence of the business cycles over time. This result was reported for the EU economies by Ormerod and Mounfield (2002) and Ormerod (2005).

However, we can look more closely within these overall results and gain further insights into business cycle synchronisation.

A graphical representation of the issue is provided by the technique of agglomerative hierarchical clustering. (Kaufman and Rousseeuw (1990)[4]). The approach constructs a hierarchy of clusters. At first, each observation is a small cluster by itself. Clusters are merged until only one large cluster remains which contains all the observations. At each stage the two 'nearest' clusters are combined to form one larger cluster. In the results presented here, the distance between two clusters is the average of the dissimilarities between the points in one cluster and the points in the other cluster[5].

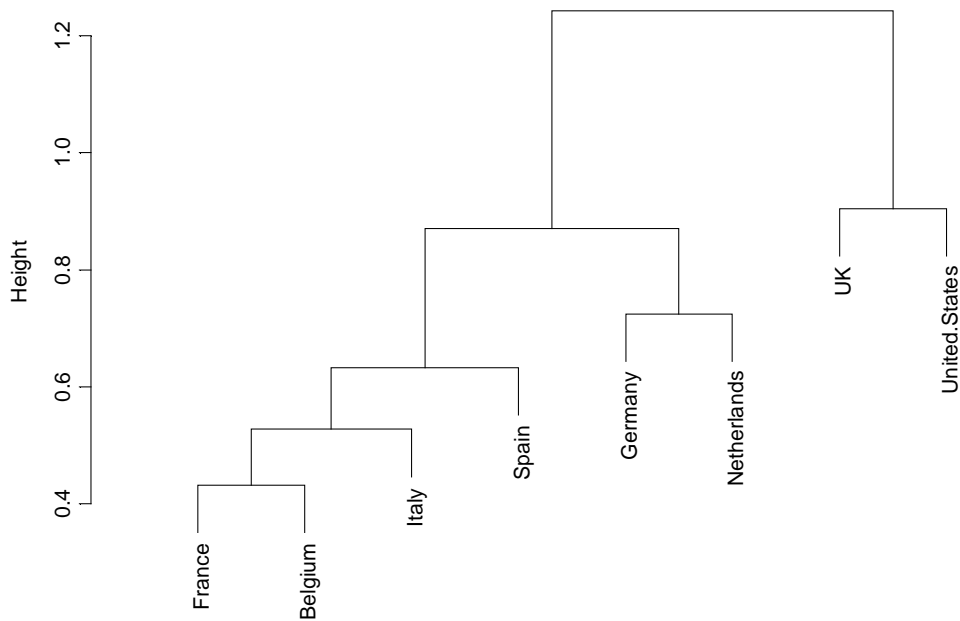

**Figure 2** *Agglomerative hierarchical clustering of the correlation matrix of annual real GDP growth rates 1981Q1-2008Q1*
.

A certain amount of exposition of the chart may be useful. The horizontal axis is of no significance to the observed structure, and relevant information is on the vertical axis. The vertical axis measures the distance at which the economies are merged into clusters. The first two economies to be merged into a cluster are France and Belgium, and the last two are the US and the UK.

The relationship between the graphical presentation and the eigenstates analysis can be illustrated as follows. The values of the eigenvector associated with the principal eigenvalue of the correlation matrix for this group of economies lie in the range 0.23 to 0.40. However, the US has a value of 0.23 and the UK 0.27, the others being in the range 0.32 to 0.40. In other words, inspection of this eigenvector reveals that it is the Anglo-Saxon economies which move more closely together and which have a relatively lower level of convergence with the EU economies. The same result can be seen visually in Figure 2, the graphical representation of the agglomerative hierarchical clustering.

---

[4] Kaufman, L. and Rousseeuw, P.J. (1990). *Finding Groups in Data: An Introduction to Cluster Analysis*. Wiley, New York
[5] The analysis was carried out using the command 'agnes' in the statistical package S-Plus, with the default options of metric = 'euclidean' and method = 'average'.

For the full sample of countries examined, the minimum value of $100\lambda_{max}/N$ is 0.38, the mean 0.60 and the maximum 0.77. Excluding the US, these figures are respectively 0.43, 0.65, 0.81. So this group of economies excluding the US exhibits a higher level of synchronisation than if the US is included. This is confirmed by a formal Kolmogorov-Smirnov test, the null hypothesis that the distribution of $100\lambda_{max}/N$ is the same both including and excluding the United States is rejected even at a p-value of 0.00.

Further excluding the UK, the minimum value of $100\lambda_{max}/N$ is 0.44, the mean 0.70 and the maximum 0.85. Again, a formal Kolmogorov-Smirnov test, the null hypothesis that the distribution of $100\lambda_{max}/N$ is the same both for the EU economies including and excluding the UK, the US being excluded from both, is rejected even at a p-value of 0.00.

Figure 3 plots the temporal evolution, suing an 8 year window again, of the main EU economies, excluding the UK and also excluding form the analysis the US.

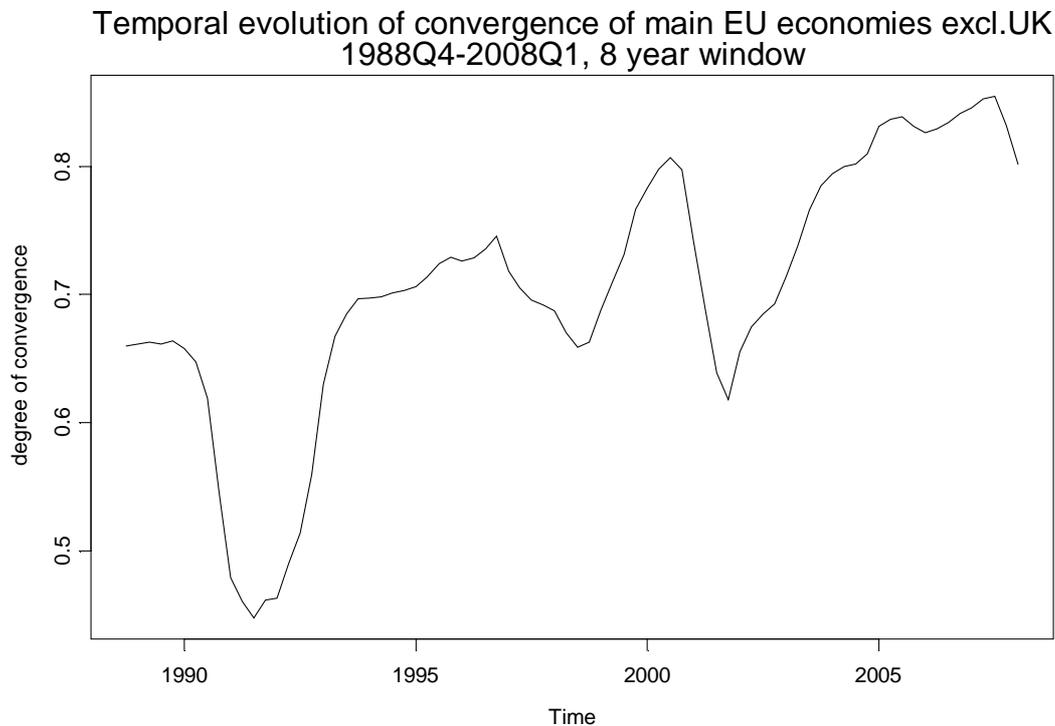

**Figure 3**   *The temporal evolution of the degree of information content in the maximum eigenvalue of the empirical correlation matrix formed from the time series of*

*quarterly GDP growth for the economies of France, Germany, Italy, Belgium, Netherlands, Spain. Each window of data spans 32 quarterly observations. The period 1981Q1 - 1988Q4 corresponds to the first data point in Figure 1, 1981Q2 - 1989Q1 to the second, and so on through to 2000Q2 - 2008Q1.*

The overall pattern is very similar to that of Figure 1. But the important difference is the range over which the data moves. The main EU economies excluding the UK have consistently exhibited a greater degree of business cycle synchronisation that if the UK is included. The UK moves in general much more in line with the US than with its EU partners.

**4.    Conclusion**

In this paper, I analyse the convergence or otherwise of the business cycle in the main economies of the European Union plus the United States, using the annual growth rates of quarterly real GDP over the 1981Q1 - 2008Q1 period. The correlations between the growth rates are analysed using random matrix theory, which enables us to identify the extent to which the correlations contain true information rather than noise.

The analysis could readily be extended to include a wider range of countries, or to examine regional convergence wither within a country or across the EU as a whole[6].

For the core EU countries, France, Germany, Italy, Spain, Belgium and the Netherlands, business cycles have shown strong synchronisation over the whole of the 1981-2008 period. The United Kingdom and the United States are considerably more synchronised with each other than they are with the main EU economies.

---

[6] As suggested to me by Pietro Terna of the University of Torino